\newif\ifJFR
\JFRfalse
\newif\ifLNCS
\LNCSfalse
\ifJFR
\documentclass{jfrarticle}
\usepackage{minted}\definecolor{vertfonce}{rgb}{0,.45,0}
\else
\ifLNCS
\documentclass{llncs}
\usepackage{minted}\definecolor{vertfonce}{rgb}{0,.45,0}
\else  
\documentclass{article}
\usepackage{color}

\usepackage{alltt}
\newenvironment{minted}[1]%
{\begin{samepage}\begin{alltt}}%
{\end{alltt}\end{samepage}}
\fi
\fi
\usepackage[all]{xy}
\usepackage{qsymbols}
\usepackage{url}

\newcommand{\Al}{\mbox{\footnotesize \textsf{A}}}
\newcommand{\Be}{\mbox{\footnotesize \textsf{B}}}

\newcommand{\Coq}{\textsf{Coq}}
\newcommand{\game}[1]{ \triangleleft\! #1\!\triangleright}

\newcommand{\nat}{\ensuremath{\mathbb{N}}}

\newcommand{\nodA}{*++[o][F]{\Al}}

\newcommand{\nodB}{*++[o][F]{\Be}}
\newcommand{\node}{*++[o][F]{~}}
\newcommand{\fl}[1]{\ar@/^/[#1]^r \ar@/^/[d]^{d}}
\newcommand{\flr}[1]{\ar@[blue]@2@/^/[#1]^{\color{blue} 2} \ar@/^/[d]^{1}}
\newcommand{\fld}[1]{\ar@/^/[#1]^2 \ar@[blue]@2@/^/[d]^{\color{blue} 1}}
\newcommand{\yiya}{\scriptstyle{\textsf{ying},\textsf{yang}}}
\newcommand{\yayi}{\scriptstyle{\textsf{yang},\textsf{ying}}}

\newcommand{\aGame}{\xymatrix{%
    &\nodA\ar@[blue][dl]\ar@[green][d]\ar@[red][r]&\nodA\ar@[blue][d]\ar@[green][dr]\ar@[red][r]& {\scriptstyle A\mapsto \textit{\scriptsize medium}, B\mapsto 0}\\
    \nodB \ar@{.>}[r]\ar@(rd,ld)[] & \nodB\ar@{.>}[dr]\ar@/_/[u] &%
    {\scriptstyle A\mapsto \textit{\scriptsize weak}, B\mapsto 2} &%
    {\scriptstyle A\mapsto \textit{\scriptsize strong}, B\mapsto 1}&\\
    &&{\scriptstyle A\mapsto \textit{\scriptsize weak}, B\mapsto 1} } }

\newcommand{\mCh}[1]{
 \ar@[blue]@/_5pc/[#1]\ar@/_4pc/[#1]%
 \ar@[blue]@/_3pc/[#1]\ar@/_2pc/@{.>}[#1]%
\ar@[green]@/_1pc/[#1]\ar@/_/[#1]%
 \ar@[green]@/^/[#1]\ar@/^1pc/@{.>}[#1]%
\ar@[red]@/^2pc/[#1]\ar@/^3pc/[#1]%
\ar@[red]@/^4pc/[#1]\ar@/^5pc/@{.>}[#1]}
\newcommand{\aMGame}{\xymatrix{%
&&&\node\ar@[blue]@/_/[dlll]\ar@/^/[dlll]%
 \ar@[blue]@/_/[ddll]\ar@/^/@{.>}[ddll]%
\ar@[green]@/_/[dddl]\ar@/^/[dddl]%
 \ar@[green]@/_/[dddr]\ar@/^/@{.>}[dddr]%
\ar@[red]@/_/[ddrr]\ar@/^/[ddrr]%
\ar@[red]@/_/[drrr]\ar@/^/@{.>}[drrr]\\
\node\mCh{ddd}&&&&&&\node\mCh{ddd}\\
&\node\mCh{ddl}&&&&\node\mCh{ddr}&\\
&&\node\mCh{dll}&&\node\mCh{drr}&&\\
 {\scriptstyle A\mapsto \textit{\scriptsize weak}, B\mapsto 1} &&&&&&
 {\scriptstyle A\mapsto \textit{\scriptsize strong}, B\mapsto 0} 
}}

\ifLNCS \else
\newtheorem{theorem}{Theorem}

\newtheorem{example}[theorem]{Example}

\fi
\usepackage{graphicx}
\newcommand{\textabstract}{
Extensive games are tools largely used in economics to describe decision processes of
a community of agents.  In this paper we propose a formal presentation based on the
proof assistant \Coq{} which focuses mostly on infinite extensive games and their
characteristics.  \Coq{} proposes a feature called ``dependent types'', which means
that the type of an object may depend on the type of its components.  For instance,
the set of choices or the set of utilities of an agent may depend on the agent
herself.  Using dependent types, we describe formally a very general class of games
and strategy profiles, which corresponds somewhat to what game theorists are used to.
We also discuss the notions of infiniteness in game theory and how this can be
precisely described.

\newcommand{\kwds}{extensive game, infinite game, sequential game,
  coinduction, Coq, proof assistant.}

\medskip

\noindent \textbf{Keywords:} \kwds}

\newcommand{\titre}{Dependent Types for Extensive Games}

\ifLNCS
\begin{document} 
\title{\titre}
\author{Pierre Lescanne}
\institute{University of Lyon, \'Ecole normale sup\'erieure de Lyon, CNRS (LIP), \\ 46 all\'ee
d'Italie, 69364 Lyon, France}

\maketitle

\noindent\rule{\textwidth}{.4pt}

\noindent\textabstract

\noindent\rule{\textwidth}{.4pt}
\else \ifJFR
\begin{document} 
\setcounter{page}{1}
\firstfoot{Journal of Formalized Reasoning  Vol.??, No.??, ??, Pages \pages.}
\runningfoot{Journal of Formalized Reasoning Vol.??, No.??, ??}
\title{\titre}
\author{Pierre Lescanne}{Pierre Lescanne\\\\
University of Lyon, \'Ecole normale sup\'erieure de Lyon, CNRS (LIP), \\ 46 all\'ee
d'Italie, 69364 Lyon, France}
\begin{abstract}
\textabstract
\end{abstract}
\maketitle
\else 
\begin{document}
\title{\titre}
\author{Pierre Lescanne\\\\
University of Lyon, \'Ecole normale sup\'erieure de Lyon, CNRS (LIP), \\ 46 all\'ee
d'Italie, 69364 Lyon, France}
\maketitle
\begin{abstract}
\textabstract
\end{abstract}
\fi %
\section{Introduction}
\label{sec:intro}

Extensive games are used in formalization of economics and in decision
processes. Rational decision is logic, but it is not exaggerated to claim that
rational decision is essentially a computational process and therefore it should be
based on computational logic, like the calculus of inductive construction of \Coq{}
and on induction.  Moreover, an adequate description of the decision process requires
the framework to be infinite. Indeed there is no reason to assume that the process is
a priori finite, since if we do so we put strong constraints on the model which
prevents some behaviors, like for instance escalation.  Beware, in the framework of
games where agents interact, we do not say that the world is infinite, but we say
that the agents \underline{believe} that the world is infinite.  Indeed, saying that
the model is finite precludes the phenomenon of escalation, and proving, in that
case, that escalation cannot exist is begging the question.  Since we require a
computational approach to infinite processes, the natural concept in modern logic is
this of coinduction as proposed
in~\cite{DBLP:journals/acta/LescanneP12,DBLP:journals/corr/abs-1112-1185,DBLP:conf/calco/Lescanne13}.
But in this paper, by using dependent types, we revise our previous works. Thus we
allow considering formal presentations of very general classes of games, for
instance, games with very general sets of choices depending on agents or very general
sets of utilities also depending on agents.  For instance, an agent may have an
infinity of choices and another may have only one choice, or two, whereas utilities
are just ordered sets, even completely trivial ones in some counterexamples, which
shows their generality.  Similarly agents may have their own sets of utility.  Agents
may prefer flowers for their colors whereas agents use their fragrances.  By very
small changes in the formalism, we may easily describe multistage games, that are
games in which agents move simultaneously at each stage.

All the formalism has been developed in
\Coq{}~\cite{barras00:_coq_proof_assis_refer_manual}.  The reader can find scripts on
GitHub at \\
\url{https://github.com/PierreLescanne/DependentTypesForExtensiveGames}.

The paper has 8 sections. The second section presents games and strategy
profiles. Section~\ref{sec:finiteness}, Section~\ref{sec:game_no_longest} and
Section~\ref{sec:esc} talk about concepts connected with finiteness.
Section~\ref{sec:infinf} considers the way infiniteness is addressed in books on game
theory. Section~\ref{sec:conc} is the conclusion.

\section{ Games and Strategy Profiles}
\label{sec:StPG}
This presentation of extensive games differs from this
of~\cite{DBLP:journals/acta/LescanneP12,DBLP:journals/corr/abs-1112-1185,DBLP:conf/calco/Lescanne13,Abramsky:arXiv1210.4537}
in the use of dependant types.  However it has connections with composition games~\cite{DBLP:journals/corr/GhaniH16,DBLP:conf/padl/HedgesOSWZ17}.
Indeed, for simplicity, in those papers, only binary
games were considered\footnote{After Vestergaard~\cite{vestergaard06:IPL} who
  introduced this concept for finite games and finite strategy profiles.}, that is
that only two choices were offered to the agents.  In this paper, using dependent
types, we can propose a more general framework.  Associated with a game, a strategy
profile is a description of the choices taken by the agents.  The formal definitions
of games and strategy profiles relies on three entities, a~set of agents written
\texttt{Agent}, a set of choices depending on an agent~\texttt{a} written
\texttt{Choice a} and a set of utilities depending on an agent \texttt{a} written
\texttt{Utility a}.  Moreover there is a preorder on \texttt{Utility a}.  In
particular, unlike most of the presentations of games, utilities need not be natural
numbers, but can be any ordered set used by the agent.  The sets of infinite games
and of infinite strategy profiles are defined coinductively and are written
respectively \textsf{Game} and \texttt{StratProf}.

\subsubsection*{Game.}
\label{sec:game}

A game which does not correspond to a terminal position and which we call a node is written \texttt{<|a,next|>} and has two arguments:
\begin{itemize}
\item  an \emph{agent} \texttt{a}, the agent whom the node belongs to,
\item a function \texttt{next} of type \texttt{Choice a $"->"$ Game}.
\end{itemize}
We call \emph{leaf} a terminal position.  A leaf consists in a function
\begin{center}
  \texttt{($`A$ a:Agent, Utility a) $"->"$ Game}
\end{center}
i. e., a function form an agent \texttt{a} to and element of \texttt{Utility a},
which is the utility assignment at
the end of the game and which is written \texttt{<|~f~|>}.  Notice that the utility
depends on the agent.  A node game is made of an agent and of a function which
returns a game given a choice. Assume that the agent is \texttt{a} and the function is
\texttt{next}, then a node game is written \texttt{<|a,next|>}.  The formal
definition of a game is given in \Coq{} by:
\begin{minted}{coq}
CoInductive Game : Set :=
| gLeaf :  (forall a:Agent, Utility a) -> Game 
| gNode : forall (a:Agent), (Choice a -> Game) -> Game.
\end{minted}
Since this defines a \emph{coinductive}, this covers finite and infinite extensive
games.
\begin{example}\label{exa:aGame}
  Here is game with \emph{choices} {\color{blue}blue}, {\color{green}
    green} and {\color{red}red} for $\mathsf{A}$ and \emph{black} and
  \emph{dotted} for $\mathsf{B}$ and $\{weak, medium, strong\}$ as
  \emph{utilities} for $\mathsf{A}$, and $\nat$ as \emph{utilities }for
  $\mathsf{B}$.

\bigskip

  \begin{centerline}
    \aGame
  \end{centerline}
\end{example}

\subsubsection*{Strategy profile.}

A strategy profile corresponds to a non terminal position.  We call it a
node and we write it \texttt{$\ll$a,c,next$\gg$}.  It has three components:
\begin{itemize}
\item an \emph{agent} \texttt{a}, the agent whom the node belongs to,
\item a \emph{choice} \texttt{c}, which is the choice taken by agent on this specific node,
\item a function \texttt{next} of type \texttt{Choice a $"->"$ StratProf}.
\end{itemize}
A strategy profile which is a terminal position is a function
\begin{center}
  \texttt{($`A$ a:Agent, Utility~a) $"->"$ Game}
\end{center}
like for games.  Indeed there is no choice. It is written
\texttt{<<f>>}.  The inductive definition in \Coq{} of a strategy profile is:
\begin{minted}{coq}
CoInductive StratProf  : Set :=
| sLeaf : (forall a:Agent, Utility a) -> StratProf
| sNode : forall (a:Agent),
            Choice a -> (Choice a -> StratProf) -> StratProf.
\end{minted}
The two main differences with the approach
of~\cite{DBLP:journals/acta/LescanneP12,DBLP:journals/corr/abs-1112-1185,DBLP:conf/calco/Lescanne13,Abramsky:arXiv1210.4537}
lie in the fact that the set of choices and the set of utilities are not
fixed (the same for all agents, namely a pair) but depend on the agent
(dependent type).  This way we can describe a larger class of games.  In
Example~\ref{exa:aGame}, we have shown a game with choices and games
actually depending on the agents.  For instance, as we will see in
Section~\ref{sec:game_no_longest}, the sets of choices can easily be
infinite.  Since the built-in \Coq{} equality is not adequate, we define
coinductively an equality on games,
\begin{minted}{coq}
CoInductive gEqual: Game -> Game -> Prop :=
| gEqualLeaf: forall f, gEqual (<| f |>) (<| f |>)
| gEqualNode: forall (a:Agent)(next next':Choice a->Game),
  (forall (c:Choice a), gEqual (next c) (next' c)) ->  
  gEqual (<|a,next|>) (<|a,next'|>).
\end{minted}
further written \texttt{==}.

\subsubsection*{Utility assignment.}

Since \Coq{} accepts only terminating functions we define the utility assignment as a
relation:
\begin{minted}{coq}
Inductive Uassign : StratProf ->  (forall a:Agent, Utility a) -> Prop :=
| UassignLeaf: forall f, Uassign (<<f>>) f
| UassignNode: forall  (a:Agent)(c:Choice a)
                       (ua: forall a',Utility a')
                       (next:Choice a -> StratProf),
    Uassign (next c) ua  -> Uassign (<<a,c,next>>) ua.
\end{minted}

We prove that \texttt{Uassign} is a functional relation, namely that

\begin{minted}{coq}
 forall s ua ua', Uassign s ua -> Uassign s ua' -> ua=ua'..
\end{minted}
Notice that for proving this property we need an inversion tactic which is somewhat
subtle when dealing with dependent
types~\cite{chlipalacpdt2011,DBLP:conf/itp/MoninS13}.\footnote{We thank Adam Chlipala
  and Jean-Fran\c{c}ois Monin for their help on this specific example.}  Moreover for all
convergent strategy profiles (i.e., for all strategy profiles of interest, see next
section) we can prove that the function is total, i.e., that there exists always a
utility assignment associated with this convergent strategy profile.

\section{Several notions associated with finiteness}
\label{sec:finiteness}

On infinite games and strategy profiles there are several predicates capturing
notions of finiteness.

\subsubsection*{Finite Games.}
A game is finite if it has a finite number of positions.  It is naturally an
inductive\footnote{Roughly speaking, an inductive (definition) is a well-founded
  definition with basic cases and constructors}.  Clearly a leaf is finite. A game
which is a node is finite if the set of the choices of the agent is
finite~\footnote{The predicate \texttt{finite} over choices is not defined here.} and
if for all the choices, the next games are finite.  This is made precise by the following
definition.

\begin{minted}{coq}
Inductive Finite : Game -> Set :=
| finGLeaf: forall f, Finite <|f|>
| finGNode: forall (a:Agent)(next: Choice a -> Game),
              finite (Choice a) ->
              (forall c:Choice a, Finite (next c)) ->
              Finite <|a,next|>.
\end{minted}

\emph{Finite strategy profiles} would be defined likewise. 
\begin{minted}{coq}
Inductive FiniteStratProf : StratProf -> Set :=
| finSLeaf: forall f, FiniteStratProf <<f>>
| finSNode: forall (a:Agent)(c:Choice a)(next: Choice a -> StratProf),
              finite (Choice a) ->
              (forall c':Choice a, FiniteStratProf (next c')) ->
              FiniteStratProf <<a,c,next>>.
\end{minted}

\subsubsection*{Games with only finitely many strategy profiles.}

Osborne and Rubinstein~\cite{osborne94:_cours_game_theory_full_first_name} call ``finite'', a game
with only finitely many strategy profiles\footnote{Actually they use the concept of
  ``history'' (path), instead of strategy profiles, but this is not essential.}.
In order not to interfere with the previous definition, we prefer to say that
\textit{the game is finitely broad}.\footnote{Denoted by the predicate
\texttt{FinitelyBroad} on \texttt{Game} in \Coq}  This is translated by the
fact that for a game \texttt{g} to have only finitely many strategy profiles, there
shall exist a list that collects all the strategy profiles that have this game
\texttt{g} as underlying game. Since in \Coq{} lists are finite this yields the
desired property:

\begin{minted}{coq}
Definition FinitelyBroad (g:Game): Prop :=
  exists (l: list StratProf), forall (s:StratProf),
      game s == g <-> In s l.
\end{minted}

\subsubsection*{Games with only finite histories.}

A game has only finite  histories if it has only finitely many paths (histories) from the root to the
leaves. This can be described as follows:

\begin{minted}{coq}
Inductive FiniteHistoryGame : Game -> Prop :=
| finHorGLeaf: forall f, FiniteHistoryGame <|f|>
| finHorGNode: forall (a:Agent)(next: Choice a -> Game),
             (forall c':Choice a, FiniteHistoryGame (next c')) ->
             FiniteHistoryGame <|a,next|>.
\end{minted}

Those games should not be confused with games with finite horizon.
Notice that Osborne and Rubinstein~\cite{osborne94:_cours_game_theory_full_first_name} require a game
with a finite horizon to have only finitely many strategy profiles (p.~90: ``[Given a
finite game] if the longest history is finite then the game has finite horizon''),
whereas Osborne~\cite{osborne04a} does not require the set of strategy profiles
associated to the game to be finite (see Section~\ref{sec:infinf}).
For strategy profiles we have:
\begin{minted}{coq}
Inductive FiniteHistoryStratProf : StratProf -> Prop :=
| finHorSLeaf: forall f, FiniteHistoryStratProf <<f>>
| finHorSNode: forall (a:Agent) (c:Choice a)
                      (next: Choice a -> StratProf),
             (forall c':Choice a, FiniteHistoryStratProf (next c')) ->
             FiniteHistoryStratProf <<a,c,next>>.
\end{minted}

\subsubsection*{Convergent strategy profiles.}

The finiteness does not apply to all paths (histories) leading to leaves,
but applies only to paths corresponding to the choices of the agents.  Mutatis
mutandi, the expression
\begin{minted}{coq}
  (forall c':Choice a, FiniteHistoryStratProf (next c'))
\end{minted}
is just replaced by
\begin{minted}{coq}
  Convergent (next c)
\end{minted}
hence without the
\begin{minted}{coq}
  forall c':Choice a
\end{minted}

Related to induction reasoning, this convergence of
strategy profiles captures continuity.  Like for the predicate
\texttt{FiniteHistoryGame} a leaf is convergent.  A~strategy profile which is a node
is convergent if the strategy subprofile for the choice made by the agent \texttt{a}
(i.e., \texttt{next c}) is convergent.

\begin{minted}{coq}
Inductive Convergent: StratProf -> Prop :=
| ConvLeaf: forall f, Convergent <<f>>
| ConvNode: forall (a:Agent) (c:Choice a)
                   (next: Choice a -> StratProf),
             Convergent (next c) ->
             Convergent <<a,c,next>>.
\end{minted}

The reader may notice the similarity of that definition with this of
finite histories for games.  We are now able to prove a theorem  on the totality of
\texttt{Uassign}:
\begin{minted}{coq}
Lemma ExistenceUassign:
  forall (s:StratProf),
    (Convergent s) -> exists (ua: forall a, Utility a), Uassign s ua.
\end{minted}
Convergence is extended to all the strategy subprofiles of a given strategy profile
by a modality \texttt{Always}, abbreviated $\Box$, when used in expressions.
\texttt{Always} applies to a predicate on \texttt{StratProf} i.e. a function
\texttt{P:StratProf $"->"$ Prop} \texttt{Always P s} means that \texttt{P} is
fulfilled by all subprofiles of \texttt{s}.

\begin{minted}{coq}
CoInductive Always (P:StratProf -> Prop) : StratProf -> Prop :=
| AlwaysLeaf : forall f, Always P (<<f>>)
| AlwaysNode : forall (a:Agent)(c:Choice a)
                      (next:Choice a->StratProf),
          P (<<a,c,next>>) ->  (forall c', Always P (next c')) ->
               Always P (<<a,c,next>>).
\end{minted}

The predicate \texttt{Always Convergent} is shortened in $\Downarrow$.
\texttt{$\Downarrow$ s} means that \texttt{s} is convergent and also all subprofiles
are convergent.  It plays a main role in the definition of other concepts related
to strategy profiles, namely equilibria and escalation.  Always convergent
strategy profiles are the right objects, that game theorists are interested in.
``Always Convergence'' captures the notion of continuity in the spirit of
Brouwer~\cite{sep-continuity}.\footnote{I like to thank Jules Hedges for pointing me
  this fact and the connection with Brouwer bar recursion~\cite{hedges16:_towar}.}

\section{A game with only finite histories and no longest history}
\label{sec:game_no_longest}

In this section we show how \Coq{} can be used to prove formally properties about
games.  Specifically we give an example of a game with only finite histories and no
longest history as a counterexample to Osborne (see~\cite{osborne04a} p.~157)
definition of finite horizon.  The game has two agents whom we call \texttt{Alice}
and \texttt{Bob} and its definition uses a feature of dependent types, namely that
the choices may depend on the agent.  In this case, \texttt{Alice} has infinitely
many choices, namely the set \texttt{nat} of natural numbers and \texttt{Bob}
has one choice, namely the set \texttt{unit}. The \emph{utility} of \emph{Alice} and
\emph{Bob} are meaningless
since they are singletons, namely the \Coq{} built-in \texttt{unit} which contains the
only element \texttt{tt}.  In \Coq{} we have:

\begin{minted}{coq}
Definition Choice :(AliceBob -> Set) :=
  fun a:AliceBob  => match a with Alice => nat | Bob => unit end.
\end{minted}
and
\begin{minted}{haskell}
Definition Utility: AliceBob -> Set := fun a => unit.
\end{minted}
Notice that \texttt{Choice} and \texttt{Utility} are functions which take an agent
and return a set. Said otherwise, the set of choices is the result of the function
\texttt{Choice} applied to agents and the set of utilities is the result of the
function \texttt{Utility} applied to agents. If the agent is \texttt{Alice}, the set
of choices is \texttt{nat} and the set of utility is \texttt{unit}.  If the agent is
\textsf{Bob} the set of choices and the set of utilities are \texttt{unit} (a
singleton).  In other words, the set of choices depends on the agents and the set of
utilities  looks depending on the agents, but doesn't.  The game has infinitely many
threadlike subgames of length $n$:

\begin{minted}{coq}
Fixpoint  ThreadlikeGame (n:nat): (Game AliceBob Choice Utility) :=
  match n with
    | 0 => <|fun (a:AliceBob) => match a with | Alice => tt
                                              | Bob => tt end|>
    | (S n) => <|Bob,fun c:Choice Bob
                     => match c with tt=>ThreadlikeGame n end|>
  end.
\end{minted}

\begin{figure}[!ht]
  \centering
\begin{displaymath}
    \xymatrix@C=15pt{
      &&*++[o][F]{\Al}\ar@{-}[lld]\ar@{-}[ld]\ar@{-}[d]\ar@{-}[rd]\ar@{.}[rrd]\ar@{.}[rrrd]\ar@{.}[rrrrd]\ar@{.}[rrrrrd]\\
      *++[o][F]{\scriptstyle \game{\,triv\,}} & *++[o][F]{\Be} \ar@{-}[d]& *++[o][F]{\Be} \ar@{-}[d]& *++[o][F]{\Be}\ar@{-}[d] &&&&& \\
      & *++[o][F]{\scriptstyle \game{\,triv\,}} & *++[o][F]{\Be} \ar@{-}[d]& *++[o][F]{\Be} \ar@{-}[d] & ......\\
      && *++[o][F]{\scriptstyle \game{\,triv\,}} & *++[o][F]{\Be} \ar@{-}[d]\\
      &&& *++[o][F]{\scriptstyle \game{\,triv\,}}
    }
  \end{displaymath}
  \caption{Picture of game with finite histories and no longest history}
  \label{fig:game}
\end{figure}
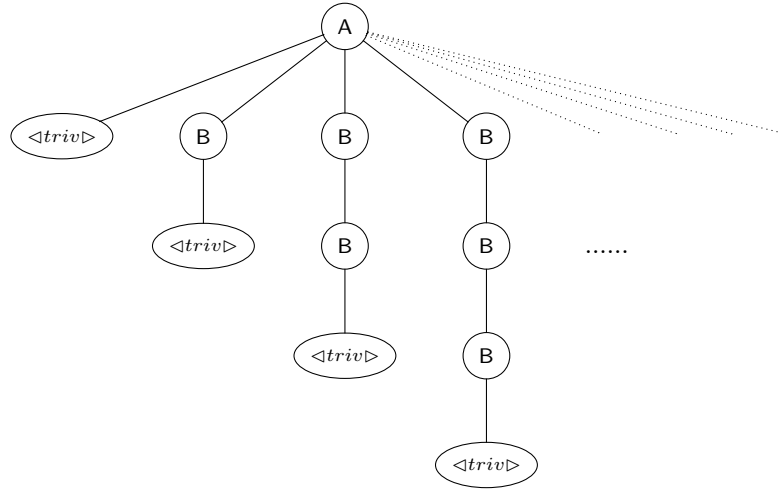

The game we are interested in is called \texttt{GameWFH} and is defined as a node
with agent
\texttt{Alice} and with next games \texttt{ThreadlikeGame n} for Alice's choice \texttt{n}: 
\begin{minted}{coq}
Definition GameWFH:(Game AliceBob Choice Utility)  :=
  <| Alice, fun n:Choice Alice => ThreadlikeGame n |>.
\end{minted}
Let us call \texttt{triv} the utility assignment \texttt{Alice => tt, Bob => tt}.  We
can picture \texttt{GameWFH} like in Figure~\ref{fig:game}.
One can prove that \texttt{ThreadlikeGame n} has only  finite histories:
\begin{minted}{coq}
Proposition FiniteHistoryGameWFH: 
    FiniteHistoryGame AliceBob Choice Utility GameWFH.
\end{minted}
Clearly \texttt{GameWFH} has no longest history.

\section{Subgame Perfect Equilibrium}
\label{sec:SPE}

An agent is rational if her strategy is based on a strategy profile which is a
subgame perfect equilibrium.  So let us present \emph{subgame perfect equilibria}.
Subgame perfect equilibria are specific strategy profiles that fulfill some ``good''
properties.  Therefore they are presented by a predicate which we call
\texttt{SPE}. In \Coq{} this is a function of type \texttt{StratProf -> Prop}.  A
strategy profile, which is a node, is a subgame perfect equilibrium if first it is
always convergent. This is necessary to be able to compute the utility assignment.
Moreover the choice of the agent is better than or equal to other choices w.r.t. to the
utility assignment and all the strategy subprofiles of this strategy profile are
themselves subgame perfect equilibria.  A leaf is a subgame perfect
equilibrium.  This can be formalized in \Coq{}:
\begin{minted}{coq}
CoInductive SPE : StratProf -> Prop :=
| SPELeaf : forall (f: forall a:Agent, Utility a), SPE <<f>>
| SPENode :  forall (a:Agent)
                    (c c':Choice a)
                    (next:Choice a->StratProf)
                    (ua ua':forall a':Agent, Utility a'),
               Always convergent <<a,c,next>> -> 
               Uassign (next c') ua' ->  Uassign (next c) ua ->
               (pref a (ua' a) (ua a)) -> SPE (next c') ->
             SPE <<a,c,next>>.
\end{minted}

\section{The simplest escalation}
\label{sec:esc}

We discussed already the rationality of escalation in infinite
games~\cite{DBLP:journals/acta/LescanneP12,DBLP:conf/calco/Lescanne13}.  For
dependent choice games, escalation is a somewhat simple concept and consists in
adjusting the types.  The simplest escalation is probably as  follows. 
It may occur in a game in which there are two agents \texttt{Alice} and \texttt{Bob},
where each agent has two choices \texttt{down} and \texttt{right} and in which there
are two non ordered utilities \texttt{ying} and \texttt{yang}.  We use \texttt{ying}
and \texttt{yang} to insist on the fact that there is no need for numbers and no need
for an actual order among the utility values.
\begin{displaymath}
  \xymatrix@C=8pt{
*++[o][F]{\Al} \ar@/^/[r]^r \ar@/^/[d]^{d} &*++[o][F]{\Be} \ar@/^/[r]^r \ar@/^/[d]^{d}
&*++[o][F]{\Al} \ar@/^/[r]^r \ar@/^/[d]^{d} &*++[o][F]{\Be} \ar@/^/[r]^r \ar@/^/[d]^{d} 
&*++[o][F]{\Al} \ar@/^/[r]^r \ar@/^/[d]^{d} &*++[o][F]{\Be} \ar@{.>}@/^/[r]^r \ar@/^/[d]^{d} 
&\ar@{.>}@/^/[r]^r \ar@{.>}@/^/[d]^{d}&\\
\yiya&\yayi&\yiya&\yayi&\yiya&\yayi&\yiya
}
\end{displaymath}
This is basically the game studied in~\cite{DBLP:conf/calco/Lescanne13}, with the
difference that the preference in \texttt{Utility = \{ying, yang\}} is just the
equality.  In other words, agents do not need to prefer one item over the other, just
a trivial preference may lead to an escalation.  The agents are like Buridan's
ass~\cite{wiki:Buridan}, they may not know what to choose and therefore go forever.
This may look strange, but as shown by the \Coq{} script, the proof is based on
exactly the same proof  technique as this of the rationality of the escalation of the
dollar auction~\cite{Shubik:1971} as shown by the two following \Coq{} statements and
proofs:\footnote{Notice that the parameters of \textsf{StratProf} are explicit!}
\ifLNCS
\begin{minted}{coq}
Lemma AlongGoodAndDivergentInDollar :
  exists (s:StratProf dollar.Agent dollar.Choice dollar.Utility),
    AlongGood dollar.Agent dollar.Choice dollar.Utility dollar.pref s 
    /\ Divergent s.
Proof.
  exists (dollarAcBc 0).
  split.
  apply AlongGoodDolAcBc.
  apply DivergenceDolAcBc.
Qed.
\end{minted}
\else
\begin{minted}{coq}
Lemma AlongGoodAndDivergentInDollar :
  exists (s:StratProf dollar.Agent dollar.Choice dollar.Utility),
    AlongGood dollar.Agent dollar.Choice dollar.Utility dollar.pref s 
    /\textbackslash Divergent s.
Proof.
  exists (dollarAcBc 0).
  split.
  apply AlongGoodDolAcBc.
  apply DivergenceDolAcBc.
Qed.
\end{minted}
\fi
and the proof of the escalation for the \emph{YingYang game}:
\ifLNCS
\begin{minted}{coq}
Lemma AlongGoodAndDivergentInYingYang :
  exists (s:StratProf yingYang.Agent yingYang.Choice yingYang.Utility),
    AlongGood yingYang.Agent yingYang.Choice yingYang.Utility yingYang.pref s 
    /\ Divergent s.
Proof.
  exists yingYangAcBc.  
  split.
  apply AlongGoodYyAcBc.
  apply DivergenceYyAcBc.
Qed.
\end{minted}
\else
\begin{minted}{coq}
Lemma AlongGoodAndDivergentInYingYang :
  exists (s:StratProf yingYang.Agent yingYang.Choice yingYang.Utility),
    AlongGood yingYang.Agent yingYang.Choice yingYang.Utility yingYang.pref s 
    /\textbackslash Divergent s.
Proof.
  exists yingYangAcBc.  
  split.
  apply AlongGoodYyAcBc.
  apply DivergenceYyAcBc.
Qed.
\end{minted}
\fi

\section{Multi-stage games}
\label{sec:multi}

Multi-stage games are introduced in~\cite{fudenberg91:_game_theor} (Section 3.2).  We
view them as games in which a node does not belong to an agent and the choices or the
moves of all the agents are simultaneous.  Let us call \texttt{MSGame} the multi-stage
games. The simultaneous or collective choice corresponds to the type:
\begin{center}
  \texttt{(forall a: Agent, Choice a) -> MSGame}
\end{center}
or written with products:
\begin{displaymath}
    \texttt{$\prod _{\mathtt{a}`:\mathtt{Agent}} \mathtt{Choice~a}$}.
\end{displaymath}
Leaves are almost unchanged. 
The function \texttt{next} is of type
\begin{displaymath}
    \texttt{next: ($\prod _{\mathtt{a}`:\mathtt{Agent}} \mathtt{Choice~a}) "->"$ MSGame}
\end{displaymath}
and a node is just the function next:
\begin{minted}{coq}
CoInductive MSGame :=
| msgLeaf: (forall a: Agent, Utility a) -> MSGame
| msgNode: ((forall a: Agent, Choice a) -> MSGame) -> MSGame.
\end{minted}

\begin{example}
  T show the complexity of multistage games, we draw a picture of a
  simple multistage game with the same choices and utilities as
  Example~\ref{exa:aGame}.

\bigskip

\begin{centerline}
  \aMGame
\end{centerline}
\end{example}

\vspace*{20pt}

\section{Infinite and infinite}
\label{sec:infinf}

In this section, we look at the way infiniteness is dealt with in textbooks on game
theory. 
\subsection*{Two views of infiniteness}

Infiniteness is discussed by Poincar\'{e} in his book \emph{Science et m\'{e}thode}
\cite{PoincareScMeth}, where he distinguishes \emph{mathematical infinite} which we
would call today \emph{potential infinite}, and \emph{actual infinite}.  Poincar\'{e}
did not believe in such an actual infinite, but today we do accept a concept of
actual infinite which is the foundation of the theory of coinduction and infinite
games.  Let us discuss these two concepts in the case of words on the alphabet $\{a,
b\}$.  $\{a,b\}^+$ represents all the (finite) words made with the letters $a$ and
$b$, like $a, b, aa, ab, ba, bb, aaa,$ $aab, aba, abb, baa, bab, bba, bbb,
\textit{etc.}$ One can also write:
\begin{eqnarray}\label{eq:star}
  \{a,b\}^+ &=& \bigcup_{n=0}^{\infty} \{a,b\}\,\{a,b\}^n.
\end{eqnarray}
$\{a, b\}^+$ is the least fixpoint of the equation:
\begin{eqnarray*}
  X &=& \{a,b\} \cup \{a,b\} X
\end{eqnarray*}
There are infinitely many such words.  This is a first kind of infinite, indeed we
can build words of all finite lengths. $\{a,b \}^{`w}$ is the set of infinite words. Each
infinite word can be seen as a function $\nat "->" \{a, b\}$.  An infinite word
represents another kind of infinite. For instance the infinite word $ababab...$ or
$(ab)^{`w}$ corresponds to the function $if~\mathsf{even}(n)~then~a ~else~b$ and is a
typical example of actual infinite.  $\{a,b\}^{`w}$ is solution of the fixpoint
equation:
\begin{eqnarray*}
  X &=& \{a,b\} X.
\end{eqnarray*}
In $\{a,b \}^{+}$ there is no infinite objects,
but only approximations, whereas in $\{a,b \}^{`w}$ there are only infinite objects.

\begin{figure}[h!]
  \centering
    \includegraphics[width=4.8cm]{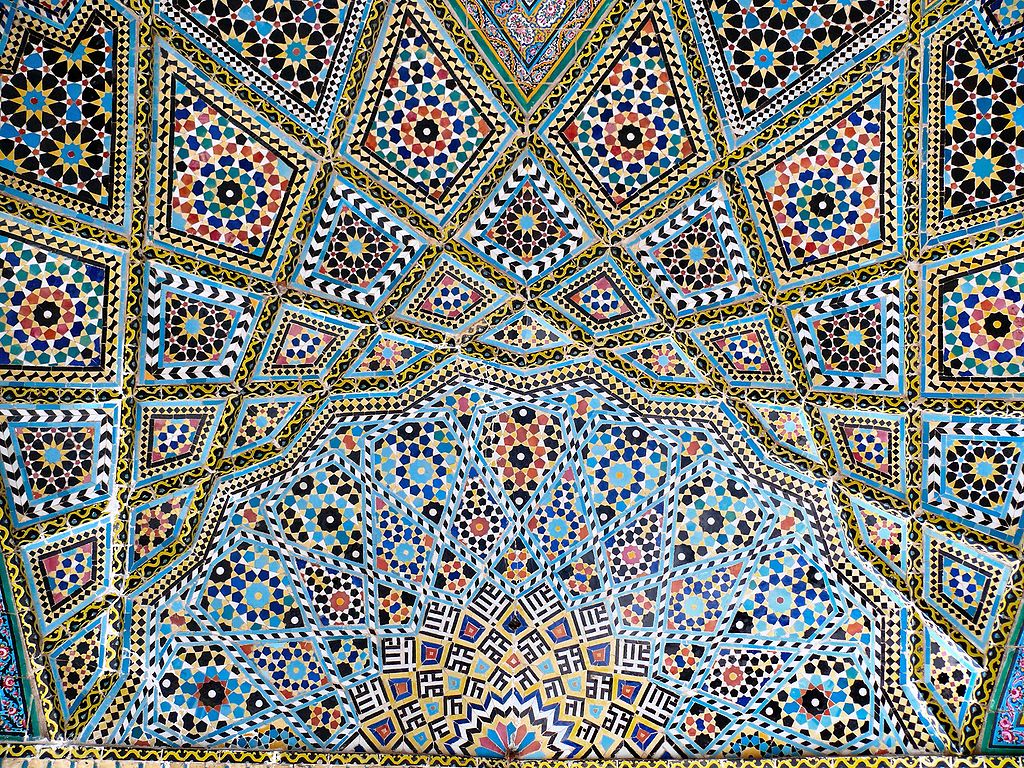} 
    \qquad\qquad\qquad\qquad
  \includegraphics[width=2.8cm]{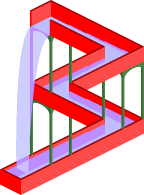}
  \caption{Two pictures of infinite}
  \label{fig:pictures}
\end{figure}

Figure~\ref{fig:pictures} represents the two notions of infiniteness. On the left, the
  vault ceiling of Nasir ol Molk Mosque in Chiraz\footnote{From Wikimedia, due to
    User:Pentocelo.} pictures
  potential infiniteness.  On the right, a drawing\footnote{From Wikimedia.} inspired by
  M.C. Escher Waterfall pictures actual infiniteness.

\subsubsection*{Common Knowledge.}

Common Knowledge is a central concept in game theory, it relies on the concept of
knowledge of an agent, which is a modality i.e., an operator of modal
logic.\footnote{We follows the presentation
  of~\cite{DBLP:journals/amai/Lescanne06,DBLP:conf/birthday/Lescanne13}, which took
  its origin from~\cite{FaginHMV95}.}  Modality $K_a$ (knowledge of agent $a$)
follows the laws of modal logic $S_5$.  For this and the group $G$ of agents, we
create a modality $E_G$ (shared knowledge):
\begin{displaymath}
  E_G(`v) = \bigwedge_{a`:G} K_a(`v).
\end{displaymath}
The common knowledge modality is 
\begin{eqnarray}\label{eq:C}
  C_G(`v) &=& \bigwedge_{n=0}^{\infty}E_G^n(`v).
\end{eqnarray}
Usually there is no ambiguity on the group of agents, thus instead $C_G$ and $E_G$
one write just $C$ and $E$.  Clearly $C$ has the flavor of $+$ as shown by the
analogy between equation~(\ref{eq:star}) and equation~(\ref{eq:C}) and their fixpoint
definitions.

\subsubsection*{Infinite and fixpoint.}

Infinite objects are associated with fixpoints.   For instance, $\{a,b\}^+$ is the
least fixpoint of the equation:
\begin{eqnarray*}
  X &=& \{a, b\} \cup \{a, b\} X
\end{eqnarray*}
whereas $C(`v)$ is the least fixpoint of 
\begin{eqnarray*}
  X  & "<=>" & `v \wedge E(X).
\end{eqnarray*}
which means that $C(`v)$ is a solution:
\begin{eqnarray*}
  C(`v) &"<=>" & `v \wedge E(C(`v)).
\end{eqnarray*}

\subsubsection*{Infinite in textbooks.}

In general in textbooks on game theory
``infinite'' is a vague notion which in not defined precisely
and words like ``ad infinitum'' (\cite{fudenberg91:_game_theor} p. 542,
\cite{shaun04:_game_theor} p.~27) or ``infinite regress''
(\cite{fudenberg91:_game_theor} p. 543) or three dots are used.  It is often said
that infinite games resemble repeated games, but this is not true, since repeated
games are typically potential infinite presentations of infinite games, i.e.,
approximation -- only sequences of games are considered, not their limit~-- whereas
infinite games are defined by coinduction.

Two main mistakes are worth  noticing.
\begin{itemize}
\item In~\cite{shaun04:_game_theor}, Hargreaves and Varoufakis  define common knowledge as
  follows:
\begin{itemize}
\item[(a)]  each person is instrumentally rational
\item[(b)] each person knows (a)
\item[(c)] each person knows (b)
\item[(d)] each person knows (c)
\item[~~] \ldots and son on \emph{ad infinitum}.
\end{itemize}
but they add ``The idea reminds one of what happens when a camera is pointing to a
television screen that conveys the image recorded by the very same camera :\emph{ an
  infinite self-reflection}'', showing that they clearly mixed up the two kinds of
notions. Indeed clearly the infinite self-reflection illustrates an actual infinite,
a little like the infinite word $(ab)^{`w}$ or the Escher waterfall, whereas, as we said,
common knowledge is a potential infinite.  An expression like \emph{ad libitum}
should have been preferred and the image of a swing going further and further or a
tessellation, like this of Figure~\ref{fig:pictures} should have been more appropriate.
\item In~\cite{osborne04a}, Osborne uses the ``length of longest terminal history''
  to define finite horizon, without checking whether this longest history actually
  exists.  A~counterexample is shown in Section~\ref{sec:game_no_longest}.  We gather
  that he means the ``least upper bound on $\overline{\nat}$ of the lengths of the
  histories''.
\end{itemize}

\section{Conclusion}
\label{sec:conc}

If, when reaching this point, the reader has the feeling that there is no proof or
almost no proof, this means that she (he) did not read the \Coq{} files of the GitHub
site, as indicated in the introduction. In those files, there is nothing but proofs.
But those proofs which are mostly meant to be read by a computer are, at the present
time, not part of a scientific paper~\cite{DBLP:journals/corr/HalesABDHHKMMNNNOPRSTTTUVZ15}.

The formalization of infinite extensive games  in \Coq{} is only at an early stage.
Among possible tracks to develop, there is the connection between multistage games
and extensive (one-stage) games, that is between games where players move simultaneously and
games where players play in alternation, using moves ``do nothing''
(see~\cite{fudenberg91:_game_theor} p.~70).   More precisely we do not know how to
intepret the sentence of Fudenberg and Tirole:
\begin{it}
  \begin{quotation}
    Common usage to the contrary ``simultaneous moves'' does not exclude games where
    players move in alternation, as we allow for the possibility that some of the
    players have the one-element choice set ``do nothing''.
  \end{quotation}
\end{it}


\end{document}

